\documentclass[pre,aps,twocolumn,superscriptaddress,citeautoscript]{revtex4-1}

\usepackage{graphicx}
\usepackage{amssymb,amsfonts,amsmath}
\usepackage{color}
\usepackage{ulem}
\usepackage[hidelinks]{hyperref}
\usepackage{bbm}
\usepackage{bbold}

\usepackage{tikz}
\usetikzlibrary{shapes}
\usetikzlibrary{patterns}
\usetikzlibrary{angles, quotes}

\newcommand{\rv}{{\mathbf r}}

\newcommand{\Tr}{{\rm Tr}\,}
\newcommand{\e}{{\rm e}}

\newcommand{\Jv}{{\bf J}}

\newcommand{\pv}{{\bf p}}

\newcommand{\fv}{{\bf f}}

\newcommand{\msphantom}[1]{$\ldots$}

\newcommand{\eqr}[1]{Eq.~\eqref{#1}}

\newcommand{\mydelete}[1]{{}}

\newcommand{\rmint}{{\rm int}}

\newcommand{\rmexc}{{\rm exc}}
\newcommand{\rmext}{{\rm ext}}

\newcommand{\rmid}{{\rm id}}

\newcommand{\cov}{{\rm cov}}

\renewcommand{\ell}{k}

\setcitestyle{super}

\begin{document}

\title{Metadensity functional learning for classical fluids:
  Regularizing with pair correlations}

\author{Stefanie M. Kampa}
\affiliation{Theoretische Physik II, Physikalisches Institut, Universit{\"a}t Bayreuth, D-95447 Bayreuth, Germany}
\author{Florian Samm\"uller}
\affiliation{Theoretische Physik II, Physikalisches Institut, Universit{\"a}t Bayreuth, D-95447 Bayreuth, Germany}
\author{Matthias Schmidt}
\email{Matthias.Schmidt@uni-bayreuth.de}
\affiliation{Theoretische Physik II, Physikalisches Institut, Universit{\"a}t Bayreuth, D-95447 Bayreuth, Germany}

\date{12 March 2026}

\begin{abstract}
   \textbf{Abstract.}
   We investigate and exploit consequences of the recent neural
   metadensity functional theory [Kampa et al.,
     \href{https://doi.org/10.1103/PhysRevLett.134.107301}{Phys.\ Rev.\ Lett.\ {\bf 134}, 107301 (2025)}] for
   describing the physics of inhomogeneous fluids. The metadensity
   dependence on the pair potential is relevant for soft matter design
   and Henderson inversion and it allows one to change the pair
   potential on the fly at prediction stage.  Here we consider
   one-dimensional systems with short-ranged (truncated) interparticle
   forces and draw on the functional pair potential dependence to
   investigate `metadirect' routes towards the bulk fluid pair
   correlation structure.  Classical density functional theory
   provides the required functional relationships.  Efficient
   variational calculus is implemented by neural functional line
   integration and automatic differentiation.  We regularize local
   learning of neural functionals by comparing the pair structure from
   different routes. Thereby results from metadirect functional
   differentiation are matched against accurate test particle data
   from an initial locally trained metadensity functional. Accessing
   the pair structure via the metadensity functional dependence
   circumvents Ornstein-Zernike inversion and it is based on first
   principles.
\end{abstract}

\maketitle

\section{Introduction}
\label{SECintroduction}

The use of machine learning has a transformative effect on the
fundamental sciences. The specific realm of soft matter research
provides a clear-cut testbed for the integration of the technology
into first-principles-based theoretical work. Classical density
functional theory \cite{hansen2013, evans1979, evans1992, evans2016,
  huang2023review} supplies suitable equilibrium cornerstones,
including the exact Mermin-Evans minimization principle
\cite{evans1979, mermin1965}, the Euler-Lagrange equation as the basis
for practical implementations, and the hierarchy of direct correlation
functionals that emerge via functional differentiation.  Associated
Ornstein-Zernike and test-particle routes give access to correlation
and response functions. A broad range of statistical mechanical sum
rules interconnect and constrain the formal objects~\cite{baus1984,
  henderson1992, hermann2021noether, mueller2024gauge,
  mueller2024whygauge, sammueller2023whatIsLiquid}.

As the classical version shares an analogous theoretical structure
with quantum density functional theory \cite{huang2023review}, it
plays arguably a similarly promising role in the age of artificial
intelligence. In particular the significant advantages in the
generation of training data for any underlying {\it classical}
particle-based system are noteworthy. A wide array of both standard
and advanced particle-based simulation methods, such as Monte Carlo
and molecular dynamics schemes, give direct access to the required
averages and, perhaps even more importantly, they can be tailored
flexibly towards the specific learning target and training
strategy~\cite{sammueller2023neural, sammueller2024attraction,
  kampa2024meta, sammueller2024hyperDFT,
  sammueller2025chemicalPotential, bui2024neuralrpm, dijkman2024ml,
  sammueller2024pairmatching, robitschko2025mixShort,
  bui2025dielectrocapillarity} (we provide an overview of further
examples below).

A significant variety of different machine learning strategies for
classical density functional theory has been put forward
\cite{cats2022ml, kelley2024ml, yatsyshin2022, malpica-morales2023,
  monti2026, fang2022, yang2024, pan2025, dijkman2024ml, ram2025ddft,
  teixera2014, lin2019ml, lin2020ml, glitsch2025disks,
  simon2024patchy, simon2023mlPatchy}. In particular the {\it local
  learning} method by Samm\"uller~{\it
  et~al.}~\cite{sammueller2023neural, sammueller2023whyNeural} has
proved to be both simple in practical use as well as highly adaptable
to a considerable array of first-principles based research goals
\cite{sammueller2023neural, sammueller2024attraction, kampa2024meta,
  sammueller2024hyperDFT, sammueller2025chemicalPotential,
  bui2024neuralrpm, sammueller2024pairmatching,
  robitschko2025mixShort}.  The method allows one to use simulation
data as an immediate basis for the accurate training of neural
functionals. In contrast to `pure' simulation work, i.e.\ computing
averages of interest via sampling under prescribed conditions, the
neural method allows one to extract constructively the {\it
  functional} relationships that are inherent in the statistical
mechanics and thus imprinted in the training data.  Functional
relationships are high-dimensional in character, as they involve maps
between function spaces. Hence traditional fitting methods are poorly
equipped for making progress. In contrast neural networks and
artificial intelligence methods are highly efficient in dealing with
high-dimensional data.

Classical density functional theory offers a wide gamut of theoretical
structure that can be exploited in practical machine learning
applications to investigate diverse physical effects within a
consistent theoretical framework. We give an overview of examples in
the following.  A locally trained hard sphere neural functional
\cite{sammueller2023neural} was shown to surpass in accuracy
Rosenfeld's fundamental measure theory for hard spheres
\cite{rosenfeld1989} in its state-of-the-art White Bear Mk~II version
\cite{roth2010}.  Thermally trained density functionals were shown to
predict accurately the liquid-gas phase behaviour of the Lennard-Jones
system~\cite{sammueller2024attraction} as well as gas-liquid and
liquid-liquid phase separation in binary mixtures
\cite{robitschko2025mixShort, zhou2026azeoptropic}.  Even very subtle
features, such as the variation of the contact angle at triple phase
coexistence, are obtained from accurate predictions of the underlying
interfacial tensions.  Bui and Cox have demonstrated the integration
of the method in their work on charged systems \cite{bui2024neuralrpm,
  bui2025dielectrocapillarity}.  Local learning gives access to neural
hyperdensity functionals that are fit for making accurate predictions
for the behaviour of general observables \cite{sammueller2024hyperDFT,
  sammueller2024whyhyperDFT}.  The method
also enables one to determine the individual values of the chemical
potential across large datasets of inhomogeneous systems and using
training data from mere canonical simulations is sufficient for local
learning~\cite{sammueller2025chemicalPotential}.

The basic mechanism of local learning is to match the output of a
neural network against simulation data for the one-body direct
correlation function $c_1(\rv)$, where $\rv$ denotes position. Results
for $c_1(\rv)$ are readily available via simple postprocessing of
sampling results for the density profile $\rho(\rv)$ obtained for a
given form of the external potential $V_\rmext(\rv)$ and prescribed
thermodynamic conditions. One then trains a neural network to either
represent the one-body direct correlation {\it functional}
$c_1(\rv;[\rho])$ or the corresponding excess (over ideal gas) free
energy functional $F_\rmexc[\rho]$ \cite{sammueller2024pairmatching};
we indicate functional dependencies by square brackets.  Both
automatic differentiation and numerical functional line integration
provide highly efficient tools to carry out flexible neural functional
calculus.  A pedagogical introduction to neural functional concepts is
given in Ref.~\citenum{sammueller2023whyNeural} using the one-dimensional
hard rod system as a simple example; a corresponding tutorial is
available online~\cite{sammueller2023neuralTutorial}.

Working with the direct correlation functionals is intimately
connected with a force-based point of view; we recall that the
one-body direct correlation functional yields the interparticle force
field as the (scaled) gradient, $k_BT\nabla c_1(\rv;[\rho])$, where
$\nabla$ denotes the derivative with respect to position $\rv$, $k_B$
is the Boltzmann constant and $T$ absolute temperature.  In {\it
  nonequilibrium} systems the interparticle force field depends in
general on time $t$ and the (causal) functional dependence is on both
the dynamcial density profile $\rho(\rv,t)$ and on the one-body
current $\Jv(\rv,t)$ \cite{schmidt2022rmp}. Hence the interparticle
force functional $\fv_\rmint(\rv,t;[\rho,\Jv])$ constitutes an
appropriate target of the neural functional
training~\cite{delasheras2023perspective, zimmermann2024ml}. Power
functional theory provides the required unique functional
relationships and an underlying exact minimization principle for the
nonequilibrium dynamics~\cite{schmidt2022rmp,
  delasheras2023perspective, zimmermann2024ml, fortini2014prl,
  krinninger2016prl, delasheras2018velocityGradient,
  stuhlmueller2018prl, hermann2019prl, delasheras2020fourForces,
  treffenstaedt2021dtpl, renner2022prl}.

As local learning poses only very moderate demands on the compute load
for producing the training data, the method was applied successfully
in equilibrium by Kampa {\it et~al.}~\cite{kampa2024meta}, in
generalization of thermal training~\cite{sammueller2024attraction,
  robitschko2025mixShort, sammueller2025chemicalPotential}, to
one-dimensional fluids interacting with {\it arbitrary} short-ranged
(truncated) pair potentials~\cite{kampa2024meta}. The neural training
thereby encompasses both the density functional dependence on
$\rho(x)$ as well as the `metadensity' functional dependence on the
form of the (scaled) pair potential $\beta\phi(r)$, where~$r$ denotes
interparticle distance and $\beta=1/(k_BT)$.  Despite the rich
heritage of analytical classical density functional research, very
little is known systematically about the dependence on $\phi(r)$
beyond the simplest mean-field approximation, which is linear in
$\phi(r)$.  For bulk fluids this approximation is analogous to the
random-phase closure that sets the bulk two-body direct correlation
function $c_2^b(r)=-\beta\phi(r)$, see e.g.~Refs.~\citenum{archer2001,
  archer2004binary, archer2017meanField, archer2013lmft} for extensive
discussion of background and applications to complex systems and
e.g.\ Refs.~\citenum{schmidt1999sfmt, schmidt2000sfmtMix,
  schmidt2000sfmtStructure, schmidt2011isometric} for work to
generalize fundamental-measure concepts beyond hard core interactions.

Having access to a concrete neural representation of the metadensity
functional for general pair interaction potentials opens up a range of
exciting research perspectives, including addressing Henderson's
inversion problem to determine uniquely the pair potential for given
form of the pair distribution function \cite{henderson1974uniqueness},
as well as to address relevant questions of modern soft matter
design~\cite{kampa2024meta, coli2022scienceAdvances,
  lindquist2016communication, shermann2020}.

An alternative to local learning is the pair correlation matching
method developed by Dijkman {\it et~al}~\cite{dijkman2024ml,
  ram2025ddft}. Their training strategy is based on using simulation
results for bulk fluids only. Specifically, results for the radial
distribution function $g(r)$ are generated (or are already known) for
a range of bulk densities $\rho(\rv) = \rho_b$ and are input into a
numerical Ornstein-Zernike inversion scheme to obtain the
corresponding bulk two-body direct correlation function
$c_2^b(r;\rho_b)$. These results serve as the reference for the
supervised training of their (convolutional) neural network that
represents the excess free energy functional $F_\rmexc[\rho]$. The
training proceeds via pair-correlation matching the second density
functional derivative of $F_\rmexc[\rho]$, evaluated at {\it constant}
bulk density, against the simulation results for $c_2^b(r;\rho_b)$.

Samm\"uller and Schmidt argue \cite{sammueller2024pairmatching} that
for {\it general} reliable density functional learning a more
exhaustive exploration of the density function space, including also
spatially inhomgenenous density profiles $\rho(\rv)$, should be
beneficial and that this in particular is accomplished by the local
learning method. They also argue that pair matching constitutes an
efficient {\it regularizer} for local
learning~\cite{sammueller2024pairmatching}, as the method operates on
the two-body correlation level. The general machine learning concept
of regularized learning refers to approaches that aim to improve model
accuracy and generalization and these are often based on mere
heuristics. In contrast, pair correlation regularization is motivated
by physical first principles and it has no adverse effect on the
primary training goal of the neural functional.

Here we consider the concept of pair correlation matching in the light
of the metadensity functional dependence on the pair potential.  In
the applications that we present, we restrict ourselves to
one-dimensional fluids that interact via short-ranged (truncated) pair
potentials that vanish beyond a fixed cutoff distance.  We find this
to be sufficient for our present conceptual investigation, as technical subtleties which arise in more general geometrical setups are avoided. However, we emphasize that our theoretical considerations are not restricted a priori to one-dimensional geometry; they rather continue to apply in higher-dimensional setups and we give a perspective on possible future applications in the outlook.  We exploit
the metadensity functional dependence on $\beta\phi(r)$ via its
`metadirect' relationship with the bulk pair distribution function
$g(r)$, as is given by functional differentiation of $F_\rmexc[\rho,
  \beta\phi]$ via Evans' corresponding exact classical relationship
\cite{evans1992}.

As a refined training method, we develop and validate a two-stage
machine learning scheme where a primary metadensity functional
\cite{kampa2024meta} is used as a bootstrapping device for the
generation of accurate neural results for the pair distribution
function. This is based on test particle density functional
minimization and it thus exploits Percus' concept \cite{percus1962} to
set the external potential equal to the pair potential; see
e.g.~Refs.~\citenum{gul2024testParticle, gul2026testParticle} for recent
work. The corresponding physical situation is that of fixing one fluid
particle at the origin and investigating the response on the
surrounding system. Beyond merely obtaining $g(r)$ as an appropriately
scaled version of the inhomogeneous density profile, thermodynamic
differentiation of the neural density functional yields results for a
corresponding locally resolved fluctuation profile. Then at the second
training stage, these results are used to regularize the metadensity
functional dependence on the pair potential. We demonstrate that the
resulting regularized metadensity functional carries markedly reduced
noise artifacts and we argue that it also has significant potential
for systematic quantitative improvement over the non-regularized
version.

The paper is organized as follows.  We start in Sec.~\ref{SECoverview}
by giving an overview of the required fundamentals of classical
density functional theory in the light of the present context.
Sec.~\ref{SECornsteinZernikeBackground} contains the essentials of the
Ornstein-Zernike relationships for the two-body density, for the local
compressibility, and for the behaviour of the local fluctuation
profiles associated with general observables.  In
Sec.~\ref{SECmetadensityFunctionalDependence} we lay out the formal
consequences of the metadensity functional dependence on the pair
potential.  Sec.~\ref{SECmetapairLearning} contains a description of
the neural functional learning method, including an overview of neural
functional learning (Sec.~\ref{SECoverviewLearning}), the description
of the initial metadensity functional training
(Sec.~\ref{SECmetapairStageOneLearning}), and the neural generation of
metadensity training data together with the details of the
second-stage pair-correlation regularization
(Sec.~\ref{SECmetapairLearningWithRegularization}).
Sec.~\ref{SECresults} contains the presentation of our results for
prototypical one-dimensional systems and in Sec.~\ref{SECconclusions}
we give conclusions.

\section{Metadensity functional theory}
\label{SECtheory}

\subsection{Classical density functional overview}
\label{SECoverview}

We first lay out several formal aspects of the metadensity functional
concept for the physics of classical fluids \cite{kampa2024meta}.  We
start with an overview of classical density functional theory
\cite{evans1979, hansen2013}, whereby we make the functional
dependence on the interparticle interaction potential explicit. We
consider systems that interact via a pair potential $\phi(r)$, where
$r$ is the interparticle distance. In general the system is exposed to
an external potential $V_\rmext(\rv)$ and the thermodynamic statepoint
is determined by the chemical potential $\mu$ and by temperature $T$.

The (thermally scaled) grand potential, when expressed as a functional
of the one-body density $\rho(\rv)$, consists of the following sum:
\begin{align}
  \beta\Omega[\rho,\beta\phi] &= \beta F_\rmid[\rho]
  + \beta F_\rmexc[\rho,\beta\phi]
  \notag\\&\quad
  + \int d\rv \rho(\rv) [\beta V_\rmext(\rv)-\beta\mu],
  \label{EQOmegaFunctional}
\end{align}
where the (thermally scaled) excess free energy functional $\beta
F_\rmext[\rho,\beta\phi]$ depends on the one-body density
profile~$\rho(\rv)$ and on the scaled interparticle interaction
potential $\beta\phi(r)$, as we have made explicit in the
notation. 

The ideal gas free energy functional is given explicitly by the
following analytical expression:
\begin{align}
  \beta F_\rmid[\rho] &=
  \int d\rv \rho(\rv)\big[\ln\big(\rho(\rv)\Lambda^d\big)-1\big],
  \label{EQmetapairFidDefinition}
\end{align}
where $\Lambda$ denotes the thermal de Broglie wavelength and $d$ is
the spatial dimensionality. The Mermin-Evans density functional
minimization principle \cite{hansen2013, mermin1965, evans1979}
ascertains that $\Omega[\rho,\beta\phi]$ is minimized in equilibrium
and hence
\begin{align}
  \frac{\delta\Omega[\rho,\beta \phi]}{\delta\rho(\rv)} 
  \Big|_{\rho=\rho_0}&= 0 \qquad \text{(min)},
  \label{EQdeltaOmegadeltaDensityZero}
\end{align}
where $\rho_0(\rv)$ denotes the equilibrium density profile.  The
corresponding equilibrium value of the grand potential, $\Omega_0$, is
obtained by evaluating the grand potential density functional
\eqref{EQOmegaFunctional} at its minimum,
\begin{align}
  \Omega_0 &= \Omega[\rho_0, \beta\phi].
\end{align}
We drop the subscript of $\rho_0(\rv)$ for convenience in the
following. The one- and two-body direct correlation functionals are
given, respectively, as the first and second density functional
derivatives of the (scaled) excess free energy functional
$F_\rmexc[\rho,\beta\phi]$ according to:
\begin{align}
  c_1(\rv;[\rho,\beta\phi]) &=
  -\frac{\delta \beta F_\rmexc[\rho, \beta\phi]}{\delta\rho(\rv)}
  \Big|_{\beta\phi},
  \label{EQconeFunctional}\\
  c_2(\rv,\rv';[\rho,\beta\phi]) &=
  \frac{\delta c_1(\rv;[\rho,\beta\phi])}{\delta\rho(\rv')}
  \Big|_{\beta\phi}.
  \label{EQctwoFunctional}
\end{align}
Here the (thermally scaled) pair potential $\beta\phi(r)$ is kept
fixed when differentiating functionally with respect to the density
profile, as is standard and also implicit in the minimization
condition \eqref{EQdeltaOmegadeltaDensityZero}.

To obtain the excess free energy functional $\beta
F_\rmexc[\rho,\beta\phi]$, starting from the functional
derivative~\eqref{EQconeFunctional} yields the inversion via
functional line integration~\cite{evans1992, sammueller2023whyNeural}:
\begin{align}
  -\beta F_\rmexc[\rho,\beta\phi] &=
  \int {\cal D}[\rho] c_1(\rv;[\rho,\beta\phi])
  \label{EQFexcAsFunctionalIntegralFormal}\\
  &= \int d\rv \rho(\rv) \int_0^1 da c_1(\rv;[a\rho,\beta\phi]).
  \label{EQFexcAsFunctionalIntegralParametric}
\end{align}
In the re-writing \eqref{EQFexcAsFunctionalIntegralParametric} we have
expressed the formal density functional line integral
\eqref{EQFexcAsFunctionalIntegralFormal} via a specific and simple
`scaling' parameterization that is straightforward to implement
numerically in practice. The scaled density profile $a\rho(\rv)$
appears as the functional argument of the one-body direct correlation
functional in the integrand of
\eqr{EQFexcAsFunctionalIntegralParametric}. The parameter range is
$0\leq a \leq 1$ according to the integration limits of the parametric
integral. The (thermally scaled) pair potential $\beta\phi(r)$ is kept
fixed when performing the functional line integral, as is consistent
with the functional derivative \eqref{EQconeFunctional}; see
Refs.~\citenum{evans1992} and~\citenum{sammueller2023whyNeural} for further background
on functional line integration methods.

When working with neural functionals, the functional line integral
\eqref{EQFexcAsFunctionalIntegralParametric} was shown to yield
accurate results e.g.\ for bulk thermodynamics
\cite{sammueller2023neural} and for interfacial free energies
\cite{sammueller2024attraction, robitschko2025mixShort}.  Furthermore
the density functional derivatives \eqref{EQconeFunctional} and
\eqref{EQctwoFunctional} can be performed efficiently via automatic
differentiation \cite{baydin2018autodiff, stierle2024autodiff}.

Finally, inserting the grand potential
decomposition~\eqref{EQOmegaFunctional} into the minimization
condition \eqref{EQdeltaOmegadeltaDensityZero} and carrying out the
occurring functional derivatives yields the Euler-Lagrange equation
\begin{align}
  c_1(\rv;[\rho,\beta\phi]) &= \ln\big(\rho(\rv)\Lambda^d\big)
  + \beta V_\rmext(\rv) - \beta\mu,
  \label{EQmetapairEL}
\end{align}
which is standard \cite{hansen2013, evans1979, evans1992, evans2016};
we set $\Lambda$ to unity in the following. As before we have made the
metadensity functional dependence on $\beta\phi(r)$ explicit in the
notation on the left hand side.

\begin{figure*}
  \includegraphics[page=1,width=0.99\textwidth]{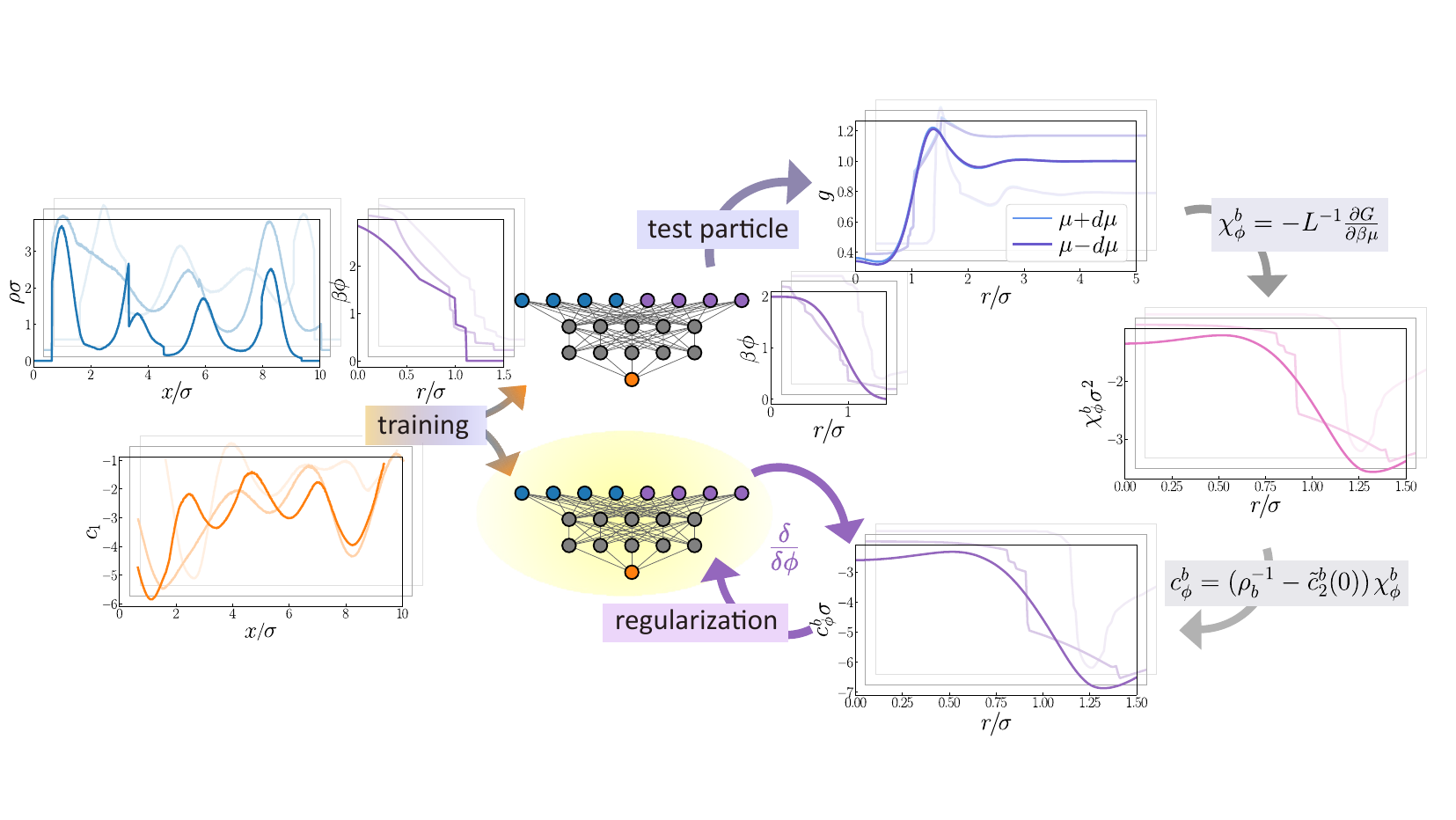}
  \caption{Overview of the two-stage neural functional learning
    (Sec.~\ref{SECmetapairLearning}).  From left to right: Training
    data consists of simulation results for inhomogeneous systems with
    density profiles $\rho(x)$ and interacting via short-ranged
    repulsive interparticle potentials $\phi(r)$
    (Sec.~\ref{SECoverviewLearning}). The corresponding one-body
    direct correlation functions $c_1(x)$ constitute the training
    target (Sec.~\ref{SECmetapairStageOneLearning}).  The initially
    trained neural functional $c_1(x;[\rho,\beta\phi])$ (top) is used
    in test particle setups for given forms of $\phi(r)$ to generate
    pair distribution functions $g(r)$
    (Sec.~\ref{SECmetapairLearningWithRegularization}).  The
    corresponding bulk metafluctuation profile $\chi_\phi^b(r) =
    -L^{-1}\partial G(r)/\partial \beta\mu$ follows efficiently via
    considering finite differences $\mu\pm d\mu$.  For the second
    training stage, the corresponding results for the bulk metadirect
    correlation function $c_\phi^b(r) = [\rho_b^{-1}-\tilde
      c_2^b(0)]\chi_\phi^b(r)$, see \eqr{EQmetapairCphiBulk}, are
    obtained. These are matched against those obtained via metadensity
    functional differentiation to regularize the second neural
    functional (bottom, highlighted),
    $c_\phi(x,r;[\rho_b,\beta\phi])=\delta
    c_1(x;[\rho_b,\beta\phi])/\delta\beta \phi(r)$, see
    \eqr{EQmetapairDefinitionCPhi}, where
    $c_\phi(x,r;[\rho_b,\beta\phi]) = c_\phi^b(r)$ is the matching
    condition \eqref{EQmetapairMatchingcphi} that is independent of
    $x$ due to the constant density profile. The second, regularized
    training stage thereby remains based also on local learning of the
    one-body direct correlation functional, as indicated.  The
    background panels illustrate further representative training
    systems (not to scale).}
\label{FIGmetapairConcept}
\end{figure*}

\subsection{Generalized Ornstein-Zernike relations}
\label{SECornsteinZernikeBackground}

We give an overview of several recent generalizations of the classical
two-body Ornstein-Zernike relation \cite{evans1979, schmidt2022rmp}.
These extensions address the local compressibility
\cite{stewart2012pre, evans2015jpcm, evans2019pnas, coe2022prl,
  coe2022pre}, the local thermal susceptibility \cite{eckert2020,
  eckert2023fluctuation, coe2022pre}, and general fluctuation profiles
that are associated with any chosen `hyperobservable' of interest
\cite{sammueller2024hyperDFT, sammueller2024whyhyperDFT}.  Three key
results are summarized: the standard inhomogeneous two-body
Ornstein-Zernike relation \cite{hansen2013, evans1979,
  schmidt2022rmp}, the local compressibility one-body Ornstein-Zernike
relation \cite{tarazona1982, eckert2020, eckert2023fluctuation}, and
the hyper-Ornstein-Zernike relation for general observables
\cite{sammueller2024hyperDFT, sammueller2024whyhyperDFT}.  The latter
case provides important background for the meta-Ornstein-Zernike
relation \cite{kampa2024meta} that we discuss below in
Sec.~\ref{SECmetadensityFunctionalDependence}. All equations of
Ornstein-Zernike-type that we summarize below follow from appropriate
differentiation of the fundamental Euler-Lagrange
equation~\eqref{EQmetapairEL}. As the respective derivatives are
consistent with the statistical mechanical structure, these equations
retain the exact nature of the Euler-Lagrange equation and hence of
the underlying fundamental minimization principle
\eqref{EQdeltaOmegadeltaDensityZero}.

The inhomogeneous Ornstein-Zernike equation for a system with one-body
density profile $\rho(\rv)$ is given by
\begin{align}
  h(\rv,\rv') &= c_2(\rv,\rv')
  +\int d\rv'' c_2(\rv,\rv'')\rho(\rv'')h(\rv'',\rv'),
  \label{EQmetapairOrnsteinZernikeInhomogeneous}
\end{align}
where the functional dependence of $c_2(\rv,\rv'';[\rho,\beta\phi])$
on the density profile and on the scaled pair potential is suppressed
in the notation.  The standard two-body total pair correlation
function is defined as $h(\rv,\rv') =
H_2(\rv,\rv')/[\rho(\rv)\rho(\rv')] - \delta(\rv-\rv')/\rho(\rv)$ with
$H_2(\rv,\rv')=\cov(\hat\rho(\rv), \hat\rho(\rv'))$, where
$\delta(\,\cdot\,)$ indicates the Dirac distribution. Here the
covariance of two phase space functions $\hat A$ and $\hat B$ is
defined in the standard way as $\cov(\hat A, \hat B) = \langle \hat A
\hat B \rangle - \langle \hat A \rangle \langle \hat B \rangle$, where
$\langle \, \cdot \, \rangle$ denotes the thermal equilibrium average,
which we specify in detail below in
Sec.~\ref{SECmetadensityFunctionalDependence}. The one-body density
operator has the standard form
$\hat\rho(\rv)=\sum_i\delta(\rv-\rv_i)$, where $\rv_i$ denotes the
position of particle $i=1,\ldots,N$.  The two-body density is then
defined as $\rho_2(\rv,\rv')=\langle
\hat\rho(\rv)\hat\rho(\rv')\rangle-\delta(\rv-\rv')\rho(\rv)$.

The local compressibility $\chi_\mu(\rv)$ is a closely related
one-body fluctuation profile \cite{tarazona1982, evans2015jpcm,
  evans2019pnas, coe2022prl, coe2022pre, coe2023, wilding2024,
  eckert2020, eckert2023fluctuation}, which is defined by
$\chi_\mu(\rv)= \beta \cov(\hat\rho(\rv),N)$, such that
$\chi_\mu(\rv)=\beta\int d\rv' H_2(\rv,\rv')$.  The corresponding
one-body fluctuation Ornstein-Zernike relation \cite{tarazona1982,
  eckert2020, eckert2023fluctuation} for the local compressibility is
given by:
\begin{align}
  \frac{\chi_\mu(\rv)}{\rho(\rv)} &= \beta +
  \int d\rv' c_2(\rv,\rv') \chi_\mu(\rv'),
  \label{EQmetapairOZcompressibility}
\end{align}
where $c_2(\rv,\rv';[\rho,\beta\phi])$ is as before the two-body
direct correlation functional.

In generalization of the local compressibility and the local thermal
susceptibility \cite{eckert2020, eckert2023fluctuation, coe2022pre},
for any given observable $\hat A$ the corresponding {\it
  hyperfluctuation} profile is $\chi_A(\rv) = \cov(\hat\rho(\rv), \hat
A)$ \cite{sammueller2024hyperDFT, sammueller2024whyhyperDFT}. The
following exact hyper-Ornstein-Zernike equation holds
\cite{sammueller2024hyperDFT, sammueller2024whyhyperDFT}:
\begin{align}
  \frac{\chi_A(\rv)}{\rho(\rv)}
    &= c_A(\rv) + \int d\rv' c_2(\rv,\rv')\chi_A(\rv),
  \label{EQmetapairHyperOZ}
\end{align}
where $c_A(\rv;[\rho,\beta\phi])$ is the one-body hyperdirect
correlation functional and in the notation its functional dependence
and that of $c_2(\rv,\rv';[\rho,\beta\phi])$ on the density profile
$\rho(\rv)$ and on the scaled pair potential $\beta\phi(r)$ are
suppressed.  We next examine closer the consequences of the latter
`metadensity' functional structure.

\subsection{Metadensity functional dependence}
\label{SECmetadensityFunctionalDependence}

Metadensity functional theory \cite{kampa2024meta} exploits the
explicit functional dependence on the scaled pair potential
$\beta\phi(r)$. We summarize several key points of the approach.  The
intrinsic free energy functional is the sum $F[\rho,\beta\phi]=F_\rmid[\rho] +
F_\rmexc[\rho,\beta\phi]$, where we recall the explicit form
\eqref{EQmetapairFidDefinition} of the ideal contribution and that
$F_\rmexc[\rho,\beta\phi]$ accounts for the interparticle interaction
effects.  The explicit Levy constrained search form \cite{levy1979,
  dwandaru2011} of the intrinsic free energy density functional is:
\begin{align}
  \beta F[\rho,\beta \phi] &=
  \min_{f\to\rho}\Tr f (\beta H_\rmint + \ln f)
  \label{EQFexcViaLevy}
\end{align}
where $\Tr\,\cdot\,$ indicates the grand ensemble trace over all
microscopic degrees of freedom and the constrained minimization is
over those trial many-body distribution functions $f$ that generate
the given one-body density profile $\rho(\rv)$
\cite{dwandaru2011}. The integrand of the Levy form
\eqref{EQFexcViaLevy} contains the intrinsic part of the Hamiltonian,
which is defined as
\begin{align}
  H_\rmint &= \sum_i \frac{\pv_i^2}{2m} + u(\rv^N),
\end{align}
where the interparticle interaction potential $u(\rv^N)$ depends on
the position coordinates $\rv_1,\ldots, \rv_N=\rv^N$ of all~$N$
particles. For the case of only pairwise contributions this energy can
be written as the following double sum:
\begin{align}
  u(\rv^N) &= \frac{1}{2}\sum_{i=1}^N\sum_{j=1,j\neq i}^N
  \phi(|\rv_i-\rv_j|).
  \label{EQurNpairwise}
\end{align}
One may define a global measure of the pair distance distribution
\cite{kampa2024meta},
\begin{align}
  \hat G(r') &= \frac{1}{2}\sum_{i=1}^N\sum_{j=1,j\neq i}^N 
  \delta(r'-|\rv_i-\rv_j|),
  \label{EQmetapairGhatDefinition}
\end{align}
which allows one to express the interparticle
potential~\eqref{EQurNpairwise} via integration over all distances
$r'$:
\begin{align}
  u(\rv^N) &= \int_0^\infty dr' \hat G(r') \phi(r').
\end{align}
The thermal mean distribution of interparticle distances is then given
as $G(r') = \langle \hat G(r') \rangle$, where the grand ensemble
average is defined as $\langle\, \cdot\, \rangle = \Tr \,\cdot\, f_0$
with $f_0$ denoting the grand ensemble equilibrium many-body
probability distribution function. For completeness we spell out the
explicit form $f_0 = \e^{-\beta(H - \mu N))}/\Xi$, with the grand
partition sum $\Xi=\Tr \e^{-\beta(H-\mu N)}$, and the full Hamiltonian
$H=H_\rmint + \sum_i V_\rmext(\rv_i)$, where $V_\rmext(\rv)$ is the
external potential and the sum is over all $N$ particles.

When starting from the inhomogeneous two-body density distribution one
can obtain $G(r')$ via integration: $G(r')=\int d\rv \int d\rv''
\rho_2(\rv,\rv'')\delta(r'-|\rv-\rv''|)/2$.  As an alternative route,
from the Levy definition~\eqref{EQFexcViaLevy} of the intrinsic
density functional one finds via functional differentiation:
\begin{align}
  G(r';[\rho,\beta\phi]) &= 
  \frac{\delta \beta
    F_\rmexc[\rho,\beta\phi]}{\delta\beta\phi(r')} \Big|_\rho
  \label{EQmetapairGofrFromFexc}  \\
  &= 
  -\frac{\delta}{\delta\beta\phi(r')}
  \int {\cal D}[\rho] c_1(\rv;[\rho,\beta\phi]) \Big|_\rho,
  \label{EQGfunctionalViaFormalIntegral}
\end{align}
where the functional line integral form
\eqref{EQGfunctionalViaFormalIntegral} follows from writing out the
excess free energy as the formal functional
integral~\eqref{EQFexcAsFunctionalIntegralFormal}. The identity
\eqref{EQmetapairGofrFromFexc} is an equivalent re-writing of Evans'
original result \cite{evans1992}.  Exchanging the orders of functional
differentiation and functional line integration in
\eqr{EQGfunctionalViaFormalIntegral} yields
\begin{align}
  G(r';[\rho,\beta\phi]) 
  &=  -\int {\cal D}[\rho] c_\phi(\rv,r';[\rho,\beta\phi])
  \notag\\
  &= -\int d\rv\rho(\rv)\int_0^1 da c_\phi(\rv,r';[a\rho,\beta\phi]),
  \label{EQmetapairGfromFunctionalIntegral}
\end{align}
where the latter form is the standard (linear scaling)
parameterization in density space. Thereby the `metadirect'
correlation functional is defined as \cite{kampa2024meta}:
\begin{align}
  c_\phi(\rv,r';[\rho,\beta\phi])
  &= \frac{\delta c_1(\rv,[\rho,\beta\phi])}{\delta\beta\phi(r')}
  \Big|_{\rho}.
  \label{EQmetapairDefinitionCPhi}
\end{align}
As is indicated in the notation, the density profile is kept fixed
upon building the functional derivative with respect to the scaled
pair potential $\beta\phi(r')$.

As a simple illustration of the metadensity functional dependence, we
consider the mean-field approximation $F_\rmexc^{\rm MF}[\rho]=\int
d\rv d\rv'\rho(\rv)\rho(\rv')\phi(|\rv-\rv'|)/2$, see
e.g.~Refs.~\citenum{archer2001, archer2004binary, archer2017meanField,
  archer2013lmft} for applications to penetrable pair potentials and
Refs.~\citenum{stewart2012pre, evans2015jpcm, evans2019pnas, coe2022prl,
  coe2022pre} for the treatment of interparticle attraction on top of
(hard core) repulsion.  The bilinear density functional dependence
leads to the one-body direct correlation functional
\eqref{EQconeFunctional} being linear in density,
$c_1(\rv;[\rho,\beta\phi]) = -\int d\rv' \rho(\rv')
\beta\phi(|\rv-\rv'|)$. The corresponding form of the metadirect
correlation functional \eqref{EQmetapairDefinitionCPhi} follows as
$c_\phi(\rv,r';[\rho])= -\int d\rv'' \rho(\rv'')
\delta(r'-|\rv-\rv''|)$. That this result is independent of
$\beta\phi(r)$ is a feature of the mean-field approximation and will
in general not hold.

One central role of the (formally exact) metadirect correlation
functional \eqref{EQmetapairDefinitionCPhi} is its appearance in the
meta-Ornstein-Zernike relation \cite{kampa2024meta}, which is exact
and given by
\begin{align}
  \frac{\chi_\phi(\rv,r')}{\rho(\rv)} &= 
  c_\phi(\rv,r')  +\int d\rv'' c_2(\rv,\rv'') \chi_\phi(\rv'',r').
  \label{EQmetaOrnsteinZernikeRelation}
\end{align}

The functional arguments of $c_\phi(\rv,r';[\rho,\beta\phi])$ and of
$c_2(\rv,\rv'';[\rho,\beta\phi])$ have again been dropped for
notational brevity. The metafluctuation profile $\chi_\phi(\rv,r')$ is
a measure of the correlations between the local density and the global
interparticle distance distribution, given by the following
covariance:
\begin{align}
  \chi_\phi(\rv,r') &= -\cov(\hat\rho(\rv),\hat G(r')).
  \label{EQmetapairChiPhiDefinition}
\end{align}

The relationship of Eqs.~\eqref{EQmetaOrnsteinZernikeRelation} and
\eqref{EQmetapairChiPhiDefinition} with the general
hyper-Ornstein-Zernike relation \eqref{EQmetapairHyperOZ} is via
choosing the general hyperobservable to be $\hat A=-\hat G(r')$, see
the definition \eqref{EQmetapairGhatDefinition} of the pair distance
observable $\hat G(r')$. This choice then leads to the exact
identification $c_A(\rv) = c_\phi(\rv,r')$ and $\chi_A(\rv) =
\chi_\phi(\rv,r')$.

It is natural to define also a global analog of the position-resolved
metafluctuation profile \eqref{EQmetapairChiPhiDefinition} in the
following form:
\begin{align}
  \chi_\phi^\circ(r') &= \int d\rv \chi_\phi(\rv,r')
  = - \cov(N,\hat G(r')),
  \label{EQmetapairChiPhiBulk}
\end{align}
which follows from integrating \eqr{EQmetapairChiPhiDefinition} over
position and noting that the total number of particles is $N=\int d\rv
\hat\rho(\rv)$. In a bulk fluid, where the density profile is
spatially constant, the metafluctuation profile is also spatially
constant: $\rho(\rv) = \rho_b \Rightarrow \chi_\phi(\rv, r') =
\chi_\phi^b(r')$.  We have the normalization relationship:
\begin{align}
  \chi_\phi^b(r') &= \chi_\phi^\circ(r')/V,
\end{align}
where $V$ is the sytem volume and for one-dimensional systems $V=L$
with system length $L$.

We next summarize several further relationships that are important for
the neural training described later.

\subsection{Consistency relationships}
\label{SECconsistency}

The global distance distribution function $G(r)$, recall the
definition \eqref{EQmetapairGhatDefinition} of the corresponding
observable $\hat G(r)$, is accessible via thermodynamic integration as
follows:
\begin{align}
  G(r;\mu) &= - \beta
  \int_{-\infty}^\mu d\mu' \int d\rv  \chi_\phi(\rv,r;\mu').
  \label{EQmetapiarGofrViaThermodynamicIntegration}
\end{align}
In the notation we have indicated the dependence of the
metafluctuation profile $\chi_\phi(\rv,r;\mu')$ on the chemical
potential $\mu'$ as the integration variable and we have dropped the
prime of the distance variable $r$ for simplicity. The dependence on
the thermodynamic variables $\mu, T$ arises from the average
\eqref{EQmetapairChiPhiDefinition}. A proof of the relationship
\eqref{EQmetapiarGofrViaThermodynamicIntegration} can be based on
differentiating both sides with respect to $\mu$, noting that
$\partial G(r;\mu)/\partial \mu = -\beta \cov(N, \hat G(r))$, as can
be seen by explicit calculation, and considering
\eqr{EQmetapairChiPhiBulk}.

As laid out in Sec.~\ref{SECmetadensityFunctionalDependence} in a bulk
fluid the metafluctuation profile is spatially constant,
$\chi_\phi(\rv,r;\mu) = \chi_\phi^b(r;\mu)$, where
\begin{align}
  \chi_\phi^b(r;\mu) &=
  -(\beta V)^{-1}\frac{\partial G(r;\mu)}{\partial \mu}
  \label{EQchiPhiBulkViaChemicalPotentialDerivative}
\end{align}
is the bulk `metacompressibility' and $V$ denotes the system volume.

By rearranging the meta-Ornstein-Zernike
relation~\eqref{EQmetaOrnsteinZernikeRelation} one obtains for a bulk
fluid:
\begin{align}
  \chi_\phi^b(r;\mu)
  &=\frac{c_\phi^b(r;[\rho_b,\beta\phi])}
  {\rho_b^{-1}-\tilde c_2^b(0)},
  \label{EQmetaOrnsteinZernikeBulk}
\end{align}
where $c_\phi^b(r;[\rho_b,\beta\phi])$ is the metadirect correlation
functional \eqref{EQmetapairDefinitionCPhi} evaluated at bulk fluid
density $\rho_b=\rm const$. Furthermore $\tilde c_2^b(0) = \int d\rv
c_2^b(|\rv|)$ denotes the position integral over the bulk two-body
direct correlation function~$c_2^b(r)$. Our notation $\tilde c_2^b(0)$
is indicative of the limit of vanishing wavevector of the
corresponding Fourier transform; see
e.g.~Ref.~\citenum{sammueller2024attraction}.

For one-dimensional bulk fluids, the relationship of $G(r)$ with the
standard pair distribution function $g(r)$ is via simple
normalization,
\begin{align}
  G(r) &= L\rho_b^2 g(r),
  \label{EQmetapairSmallBigG}
\end{align}
and for higher-dimensional systems appropriate geometric normalization
is required.

For general inhomogeneous states performing the integral of $G(r)$
over all distances $r$ yields, from the
definition~\eqref{EQmetapairGhatDefinition}, the following thermal
average:
\begin{align}
  \int_0^\infty dr G(r) &= \langle N(N-1)/2\rangle,
  \label{EQmetapairGofrIntegral}
\end{align}
which forms a useful sum rule to carry out consistency tests on data;
in practice the upper integration limit $\infty$ is understood to
indicate the integral over all distances that occur in the (finite)
system volume.

The present theoretical framework sets the formal constraints and
checks that are useful to set up and control the machine learning
procedures, as described in the following.

\begin{figure*}[tb]
  \includegraphics[page=1,width=0.8\textwidth]{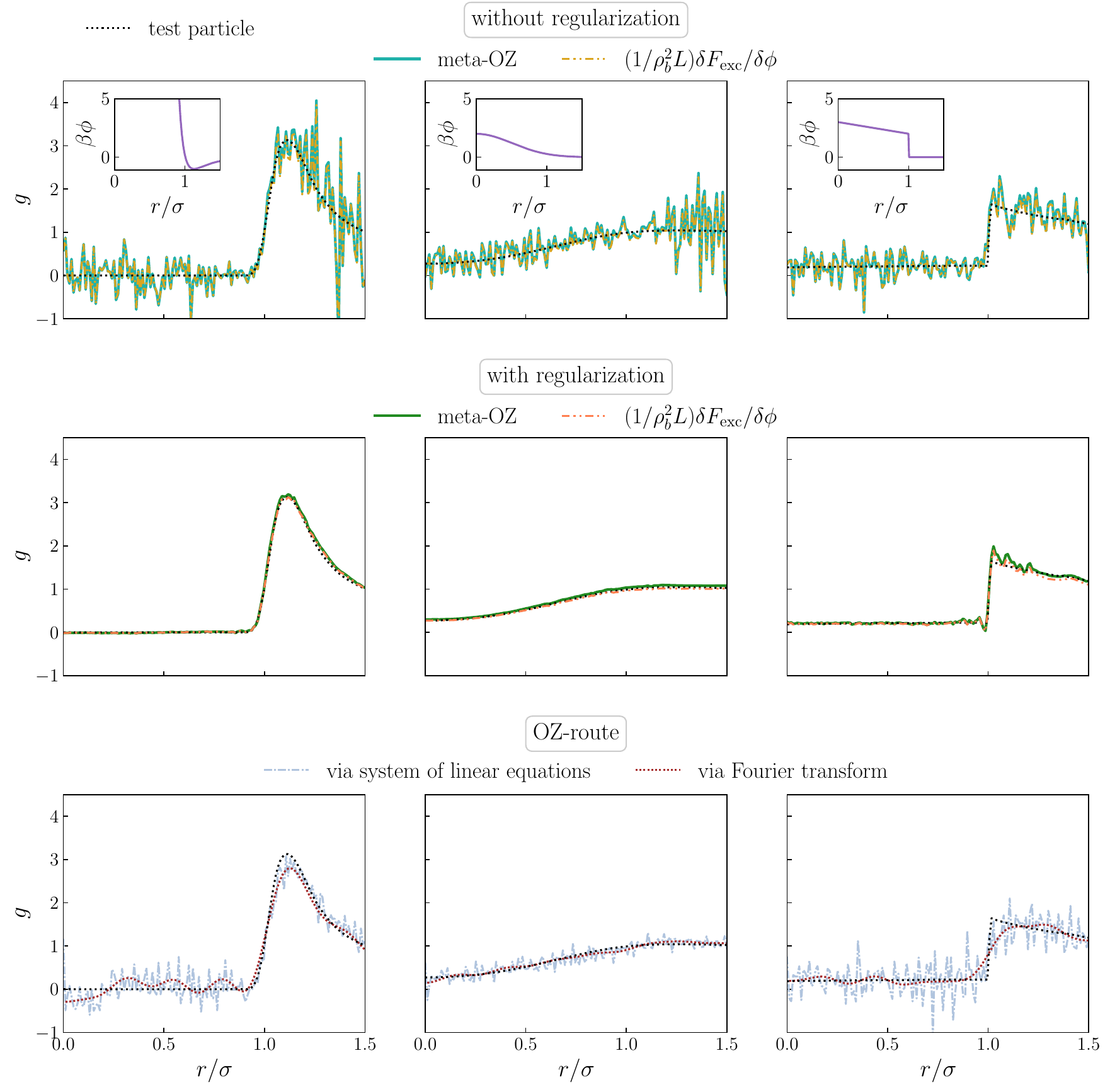}
  \caption{Neural functional results for the bulk pair distribution
    function $g(r)$ from different theoretical routes and from both
    unregularized and regularized metadensity functionals. The
    generality of both metadensity functionals is exemplified by their
    application to three different types of pair potentials (insets):
    truncated Lennard-Jones-like pair interaction (left column),
    repulsive Gaussian pair potential (middle column), and a
    penetrable-ramp potential (right column). The results for $g(r)$
    are shown for scaled distances $r/\sigma$ inside the pair
    potential window $r\leq r_c = 1.5\sigma$; the scaled chemical
    potential is $\beta\mu = 1$.  In each panel the result from
    Percus' test particle minimization (solid line) serves as the
    reference; this route was found previously to yield excellent
    agreement with simulation data \cite{kampa2024meta}.  The results
    in the first row stem from a locally-trained unregularized neural
    metadensity functional using either the meta-Ornstein-Zernike
    route [Eqs.~\eqref{EQmetapiarGofrViaThermodynamicIntegration},
      \eqref{EQmetaOrnsteinZernikeBulk}, and
      \eqref{EQmetapairSmallBigG}] or functional differentiation
    according to Eqs.~\eqref{EQGfunctionalViaFormalIntegral} and
    \eqref{EQmetapairSmallBigG}: $(\rho_b^2 L)^{-1}\delta \beta
    F_\rmexc[\rho,\beta\phi]/\delta\beta\phi(r)$ is evaluated numerically
    with stepwidth $d(\beta\phi) = 0.1$; see text.  For each system both
    unregularized metadirect results exhibit irregular deviations from
    the (test particle) reference for~$g(r)$, in particular for
    distances close to $r_c$ and at the maximum of $g(r)$.  The
    results in the second row are obtained using the pair-regularized
    neural metadensity functional following the same (metadirect)
    routes.  Noise artifacts are eliminated and the agreement with the
    (test particle) reference is excellent.  The results in the third
    row follow from using the standard density functional dependence
    instead of the metadensity channel of the neural functional.
    Numerical solution of the standard Ornstein-Zernike equation
    \eqref{EQmetapairOrnsteinZernikeInhomogeneous}, using as input the
    bulk limit of \eqr{EQctwoFunctional}, is performed either via
    solution of the corresponding set of linear equations, which
    arises upon spatial discretization of
    Eq.~\eqref{EQmetapairOrnsteinZernikeInhomogeneous}, or via spatial
    Fourier transform.  While both routes predict qualitatively
    correct behaviour, each one is prone to artificial numerical
    oscillation in regions of large scaled pair potential
    $\beta\phi(r)$ and to deviations near the first peak of $g(r)$.}
  \label{FIGmetapairThreeRoutes}
\end{figure*}

\begin{figure*}[tb]
  \includegraphics[page=1,width=0.99\textwidth]{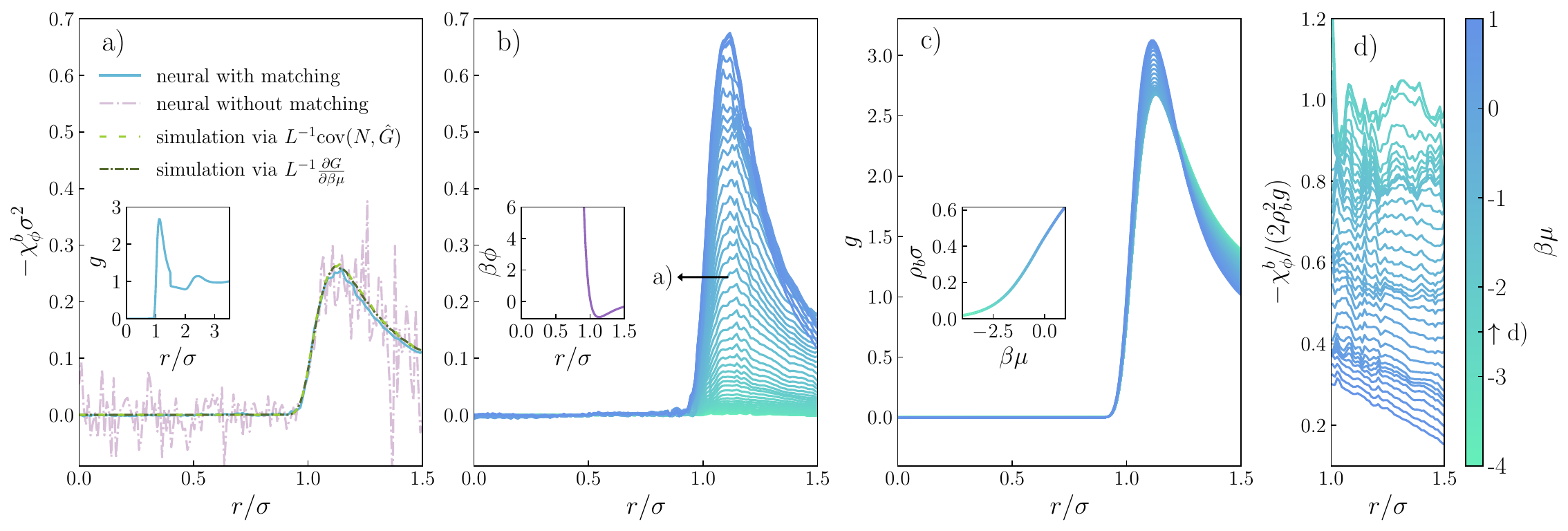}
  \caption{Illustration of the relationship of the bulk
    metacompressibility and the pair distribution function.  a)~Scaled
    bulk metacompressibility, $-\chi_\phi^b(r)\sigma^2$, of a
    one-dimensional Lennard-Jones-like fluid at scaled chemical
    potential $\beta\mu = -1$ and pair distribution function $g(r)$
    (inset).  Results for $-\chi_\phi^b (r)\sigma^2$ are shown as a
    function of the scaled distance $r/\sigma$, obtained by solving
    the bulk meta-Ornstein-Zernike equation
    \eqref{EQmetaOrnsteinZernikeBulk} using either the neural
    metadensity functional without or with regularization.  Reference
    simulation data for system size $L=5\sigma$ are obtained
    consistently by alternative methods: $-L^{-1}\partial G/\partial
    \beta \mu$ obtained from numerical differentiation of simulation
    results for $G(r)$ and spatial integration over sampled density
    covariance data according to $\chi_\phi^b(r) = -\int
    dx\cov(\hat\rho(x),\hat G(r))/L$, which is identical to sampling
    the covariance $-\cov(N, \hat G(r))/L$. The expected equivalence
    of the results is reflected by their numerical agreement.
    b)~Variation of the scaled metafluctuation profile
    $-\chi_\phi^b(r) \sigma^2$, shown as a function of scaled
    interparticle distance $r/\sigma$, over a range of different
    values of the scaled chemical potential $\beta\mu$ (colour bar).
    c)~Corresponding sequence of results for the pair distribution
    function $g(r)$, obtained via the test particle method. The inset
    shows the scaled bulk density $\rho_b\sigma$ as a function of
    $\beta\mu$ (inset).  d)~Scaled ratio
    $\chi_\phi^b(r)/\big(2\rho_b^2g(r)\big)$ shown for distances
    outside the core, $1<r/\sigma<1.5$, and for $\beta\mu>-2.5$
    (indicated by the vertical arrow at the colourbar). The low
    density limit of unity can be identified despite some noise
    artifacts and the thus scaled fluctuations are suppressed upon
    increasing bulk density.}
\label{FIGmetapairBulkMetacompressibility}
\end{figure*}

\begin{figure}[tb]
  \includegraphics[page=1,width=0.99\columnwidth]{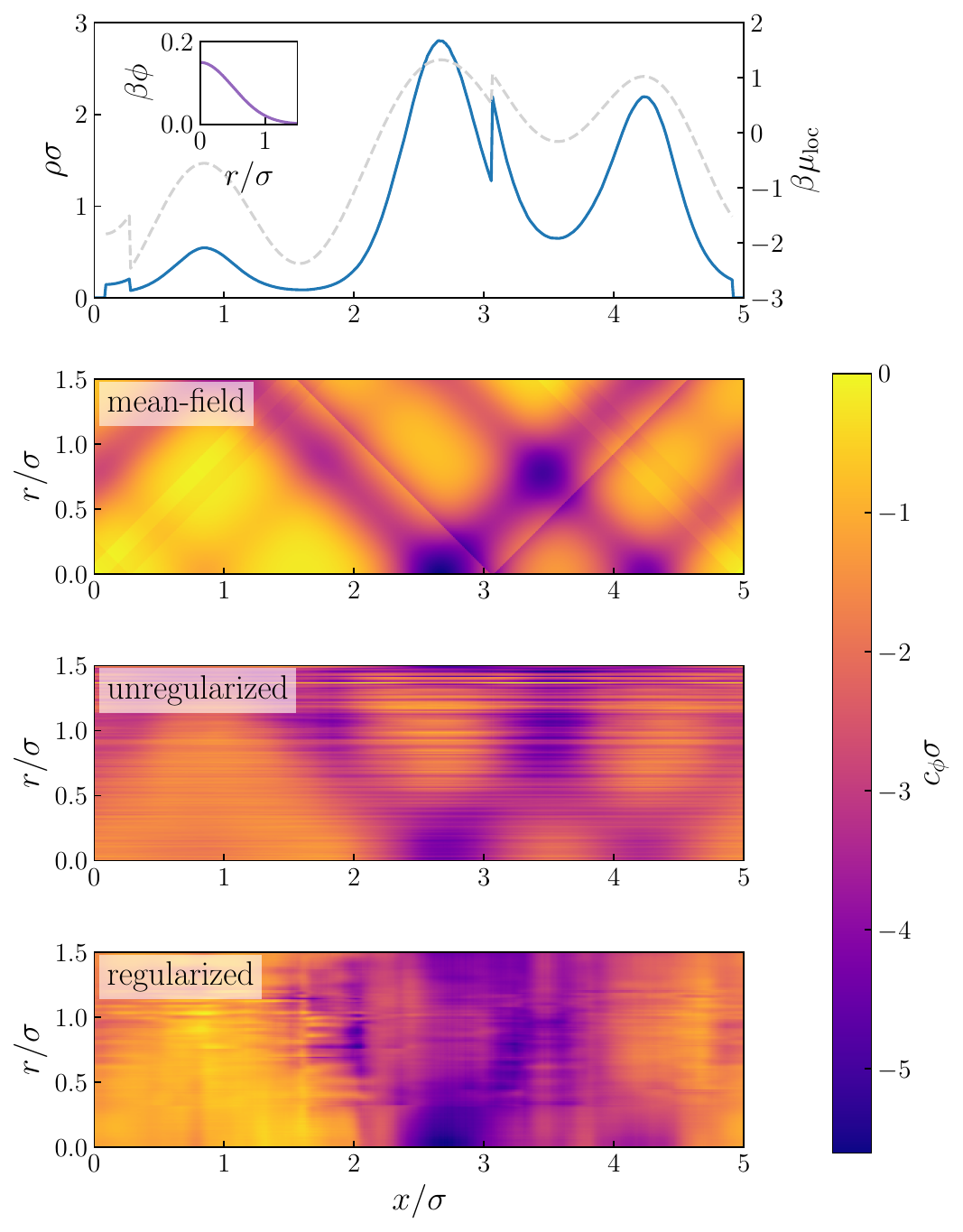}
  \caption{Illustration of the metadensity functional dependence in an
    inhomogeneous fluid with local chemical potential $\beta\mu_{\rm
      loc}(x)=\beta\mu-\beta V_\rmext(x)$ and highly structured
    density profile $\rho(x)\sigma$ (top panel). The particles
    interact mutually with a repulsive penetrable pair potential
    (inset).  The corresponding (scaled) metadirect correlation
    function $c_\phi(x,r)\sigma$, see \eqr{EQmetapairDefinitionCPhi},
    is shown as a function of (scaled) position $x/\sigma$ and
    (scaled) interparticle distance $r/\sigma$. Three different
    functionals are used to generate results: the analytic mean-field
    approximation $\int dx'\rho(x')\delta(|x'-x|-r)\sigma$ (second
    panel), as well as the unregularized (third panel) and regularized
    (bottom panel) neural one-body direct correlation functionals,
    which yield $c_\phi(x,r)$ via automatic differentiation according
    to \eqr{EQmetapairDefinitionCPhi}.}
\label{FIGmetapairMetaDirectCorrelationFunctional}
\end{figure}

\begin{figure*}[tb]
  \includegraphics[page=1,width=0.89\textwidth]{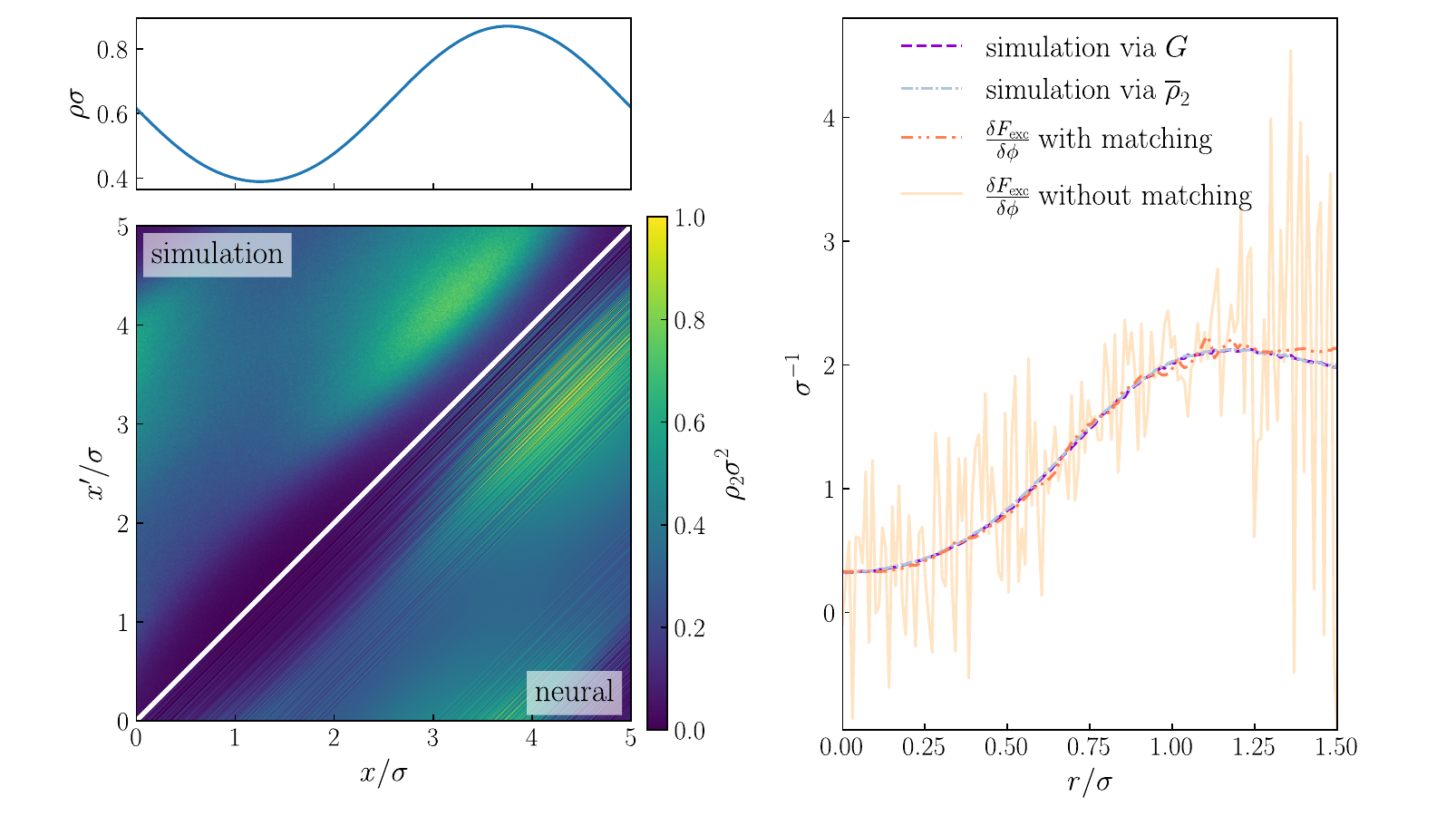}
  \caption{Demonstration of metadensity functional application in an
    inhomogeneous system. The particles interact via a repulsive
    Gaussian pair potential. The (scaled) one-body density profile
    $\rho(x)\sigma$ is sinusoidal (top left panel).  Results for the
    scaled two-particle density $\rho_2(x,x')\sigma^2$ are shown as a
    function of $x/\sigma$ and $x'/\sigma$ and indicate structure
    formation (lower left panel).  Simulation data is shown in the
    upper left triangle; the results in the lower right triangle are
    obtained from solving the inhomogeneous Ornstein-Zernike equation
    \eqref{EQmetapairOrnsteinZernikeInhomogeneous}, using neural input
    for $c_2(x,x')$, and the consistency is reflected by the mirror
    symmetry around the counter-diagonal, $\rho_2(x,x')=\rho_2(x',x)$.
    The simulation results for $G(r)$ (right panel) are obtained
    either from sampling the interparticle distance histogram or,
    equivalently, by spatial integration $\bar\rho_2(r) = \int
    dx\rho_2(x-r,x)$ with implied periodic continuation.  The
    corresponding neural results are obtained via
    \eqr{EQGfunctionalViaFormalIntegral} using the unregularized or
    the regularized density functional.}
\label{FIGmetapairTwoBodyDensity}
\end{figure*}

\section{Metadensity functional learning}
\label{SECmetapairLearning}

\subsection{Overview of local functional learning}
\label{SECoverviewLearning}

We describe in the following how the theoretical metadensity
functional structure (Sec.~\ref{SECtheory}) can be exploited to
significant effect when using simulation-based supervised machine
learning to represent the functional relationships. The neural
methodology rests on the straightforward availability of data for the
direct correlation functions via simulations for a range of specific
training systems. The training systems are constructed in a way that is
as diverse as possible at the first learning stage, which in practice
includes randomized thermodynamic conditions, randomized external
potentials, and randomized pair potentials \cite{kampa2024meta}.

Extracting the universal functional relationships from the training
data is based on a neural network topology that acts as a mere
universal approximator without any pre-configured internal
structure. This is accomplished by using a standard fully-connected
multilayer perceptron. While we find this learning setup to be optimal
for the current conceptual purposes, we see much potential for future
progress in both tailoring the training data and the neural functional
\cite{sammueller2024pairmatching}.

We give a graphical overview of the supervised machine learning
concept in Fig.~\ref{FIGmetapairConcept}. The essential steps are the
initial training of a neural metadensity functional
(Sec.~\ref{SECmetapairStageOneLearning}) and its use in the generation
of {\it neural} training data which forms the basis for the second
stage regularized
training~(Sec.~\ref{SECmetapairLearningWithRegularization}).

\subsection{Initial metadensity functional training}
\label{SECmetapairStageOneLearning}

At the first machine learning stage we use the setup described in
Ref.~\citenum{kampa2024meta} to train a neural one-body direct
correlation functional $c_1(x;[\rho,\beta\phi])$ that makes the
functional dependence on $\beta\phi(r)$ operational in practice.  The
method requires one to prepare a range of randomized training systems,
which we identify by the index $\ell$, and to sample the corresponding
density profile $\rho_\ell(x)$. Each training system $\ell$ is
specified by a corresponding randomized form of the external potential
$V_{\rmext,\ell}(x)$, randomized values of the thermodynamical
parameters $\mu_\ell$ and $\beta_\ell$, as well as a randomized form
of (truncated) scaled pair potential $\beta_\ell\phi_\ell(r)$. (In
practice, inverse temperature can be regarded as a constant scaling
parameter.)  A simple post-processing step, based on the
Euler-Lagrange equation~\eqref{EQmetapairEL}, then yields simulation
results for the one-body direct correlation function $c_{1,\ell}(x)$
for each training system as follows:
\begin{align}
  c_{1,\ell}(x) &= \ln\rho_\ell(x) 
  + \beta_\ell V_{\rmext,\ell}(x) - \beta_\ell \mu_\ell.
  \label{EQmetadirectc1fromSimulation}
\end{align}

The local learning method aims to train a neural functional
$c_1(x;[\rho,\beta\phi])$ such that equality between neural
predictions and simulation results is achieved:
\begin{align}
  c_1(x;[\rho_\ell, \beta_\ell\phi_\ell]) &= c_{1,\ell}(x),
  \quad \forall (x, \ell).
  \label{EQmetadirectc1matching}
\end{align}
The condition \eqref{EQmetadirectc1matching} encapsulates the training
goal to have the ouput of the neural functional (left hand side)
reproduce the `target' simulation data (right hand side). The matching
should hold for all positions $x$ in each training system $\ell$. As
we restrict ourselves to interparticle potentials of finite range, we
use a truncated density input `window' to represent the functional
dependence of the neural one-body direct correlation functional
\cite{kampa2024meta, sammueller2024attraction, sammueller2023neural};
we recall that the method remains useful when combined with insights
\cite{cox2020pnas, bui2024} to treat additional long-ranged
interparticle forces \cite{bui2024neuralrpm,
  bui2025dielectrocapillarity, zhou2026azeoptropic}.

\subsection{Regularized metadensity functional learning}
\label{SECmetapairLearningWithRegularization}

Equation~\eqref{EQmetadirectc1matching} is a sufficient condition for learning the functional dependence of $c_1(x)$ on both the density profile $\rho(x)$ and the scaled pair potential $\beta \phi(r)$, provided that the underlying functional mapping is adequately explored by the training data.
In practice, however, it is useful to consider additional matching conditions both to alleviate the burden of generating a sufficient amount of training data and to improve the numerical quality of the neural network predictions.
In the following, we lay out a workflow for incorporating radial distribution functions obtained from an initial (`unregularized') neural functional in the training of a second (`regularized') neural functional with the goal of improving its quality.
The idea is simple:
As metadensity functional theory facilitates multiple routes towards the pair correlation structure, enforcing consistency between different routes can be utilized as an additional matching condition for neural functional training.
Ultimately, Eq.~\eqref{EQmetapairMatchingcphi} will constitute this second matching condition and we lay out its first-principles origin in the following.

The unregularized neural metadensity functional
(Sec.~\ref{SECmetapairStageOneLearning}) is used to generate
additional (`synthetic') training data for the following regularized
learning. The generation of the training data proceeds via standard
neural density functional minimization in test particle situations via
numerical Picard iteration of the Euler-Lagrange
equation~\eqref{EQmetapairEL}. Test particle minimization is known to
be an accurate method that is broadly applicable in a variety of
contexts, including machine-learning \cite{glitsch2025disks},
fundamental-measure theory~\cite{cats2022ml, gul2024testParticle},
mean-field theory \cite{monchojorda2023}, as well as for the dynamics
\cite{archer2007dtpl, hopkins2010dtpl, brader2015dtpl,
  schindler2016dynamicPairCorrelations, treffenstaedt2021dtpl,
  treffenstaedt2022dtpl, schmidt2022rmp, stopper2019dtpl} and quantum
physics \cite{mccary2020prlBlueElectron, pederson2022conditionalDFT}.

The test particle application of the neural density functional theory
yields results for pair correlation functions $g(r)$ for given forms
of scaled pair potential $\beta_{}\phi_{}(r)$ as follows. In the
Euler-Lagrange equation \eqref{EQmetapairEL} one sets the external
potential equal to the pair potential, $V_\rmext(r)=\phi(r)$, and
hence obtains:
\begin{align}
  \rho_{g}(r) &= 
  \exp\big(c_1(r;[\rho_{g},\beta\phi])
  -\beta\phi(r)+\beta\mu\big),
  \label{EQmetapairELTestParticle}
\end{align}
which is a self-consistency equation that determines the density
profile $\rho_g(r)$ in the test particle geometry. The corresponding
bulk density $\rho_{b}=\rm const$ can be obtained by self-consistent
solution of the Euler-Lagrange equation~\eqref{EQmetapairEL} in
absence of an external potential, $V_\rmext(x)=0$, which implies:
\begin{align}
  \rho_{b} &=
  \exp\big(c_1(r;[\rho_{b},\beta\phi])+\beta\mu
  \big).
  \label{EQmetapairBulkDensity}
\end{align}
Results for the pair distribution function $g(r)$ then follow by
simple normalization:
\begin{align}
  g(r) &= \frac{\rho_g(r)}{\rho_b}.
\end{align}
Alternatively, one may obtain the bulk density $\rho_b$ by using the
limiting behaviour of the test particle density profile at large
distances, $\rho_g(r)\to\rho_b$ for $r\to\infty$.

The high numerical accuracy that is attainable in the neural test
particle minimization enables one to also obtain results for the bulk
pair metafluctuation profile $\chi_\phi^b(r)$, as evaluated via the
thermodynamical parametric derivative
\eqref{EQchiPhiBulkViaChemicalPotentialDerivative} with the required
results for $G(r)$ following from the rescaling
\eqref{EQmetapairSmallBigG}. Hence
\begin{align}
  \chi_\phi^b(r;\mu) &=
  -\beta^{-1}\frac{\partial }{\partial \mu} \rho_b^2  g(r;\mu),
  \label{EQchiBulkAsDerivative}
\end{align}
where the bulk density $\rho_b$ depends on the value of the chemical
potential $\mu$, which needs to be taken into account differentiating
parametrically with respect to~$\mu$.
Then from \eqr{EQmetaOrnsteinZernikeBulk} one obtains the
corresponding metadirect correlation function as
\begin{align}
  c_\phi^b(r) &= 
  \big[ \rho_b^{-1}-\tilde c_2^b(0) \big]
  \chi_\phi^b(r).
  \label{EQmetapairCphiBulk}
\end{align}

In total,
Eqs.~\eqref{EQmetapairBulkDensity}-\eqref{EQmetapairCphiBulk} offer a
fast and accurate means to obtain results for metadirect correlation
functions $c_\phi^b(r)$ utilizing results from the test particle
minimization~\eqref{EQmetapairELTestParticle}. Crucially, there are no
simulations employed at this stage and the results are rather solely
based on the (unregularized) neural functional described in
Sec.~\ref{SECmetapairStageOneLearning}. Its application to a range of
randomized thermodynamic parameters $\mu_k, \beta_k$ and randomized
(scaled) pair potentials $\beta_k \phi_k(r)$ hence generates an
additional (synthetic) data set $c^b_{\phi,k}(r)$ to be used for metadirect
regularization in a second training stage, as described in the
following.

At the second training stage, our setup takes the concept of pair
correlation matching \cite{dijkman2024ml, ram2025ddft}, which we
recall was shown to be an efficient density functional regularizer
\cite{sammueller2024pairmatching}, and generalizes this for
regularization of neural {\it metadensity} functionals. We thereby
exploit the avenues offered by the explicit functional dependence on
the scaled pair potential. Hence we use the following matching
condition:
\begin{align}
  c_\phi(x,r;[\rho_{b,\ell},\beta_\ell\phi_\ell]) &= c_{\phi,\ell}^b(r),
  \label{EQmetapairMatchingcphi}
\end{align}
where the left hand side is the output of the neural metadirect
correlation functional, obtained via automatic differentiation
\eqref{EQmetapairDefinitionCPhi}. The right hand side is the reference
data obtained self-consistently, starting from test particle results
for $g(r)$, which are converted via \eqr{EQchiBulkAsDerivative} to
$\chi_\phi^b(r)$, which yield via \eqr{EQmetapairCphiBulk} results for
$c_\phi^b(r)$. Thereby the neural functional results for pair
correlation functions $g(r)$ follow from test particle minimization
using the purely locally trained neural functional as a bootstrapping
device. In our notation we have retained in
\eqr{EQmetapairMatchingcphi} the dependence on position $x$ on the
left hand side; however, due to the density profile being spatially
constant the neural functional output does not depend on the value of
$x$.

\section{Results}
\label{SECresults}

We demonstrate numerically in the following that the regularized
learning method described in Sec.~\ref{SECmetapairLearning} improves
upon the `bare' local metadensity functional learning
\cite{kampa2024meta} in terms of increased accuracy of the density
functional results obtained via the `metachannel'.  We recall the
overview of our machine learning strategy displayed in
Fig.~\ref{FIGmetapairConcept}, where real data were already shown.

In Fig.~\ref{FIGmetapairThreeRoutes} we display results for the bulk
pair distribution function $g(r)$, as obtained via several different
routes and for three different representative types of short-ranged
pair potentials, namely a Lennard-Jones-like pair interaction, a
repulsive Gaussian potential, and a penetrable-ramp form.
The reference results are obtained from test particle minimization, as
was shown previously to be numerically highly consistent with
simulation data \cite{kampa2024meta}. Having concrete access to the
full metadensity functional dependence allows us to compare the
results from a range of different routes to $g(r)$, including
thermodynamic integration via
\eqr{EQmetapiarGofrViaThermodynamicIntegration} using either the
unregularized or the regularized neural functional. The respective
integrand is obtained via the bulk meta-Ornstein-Zernike equation
\eqref{EQmetaOrnsteinZernikeBulk} with all expressions on the
right-hand side being evaluated via the neural metadensity
functional. The results from the initial neural functional show
qualitatively correct features, but they also display significant
noise. In quite striking contrast, the results from the regularized
functional show excellent agreement with the test particle reference
(compare the first and second rows in
Fig.~\ref{FIGmetapairThreeRoutes}). Furthermore calculating
$F_\rmexc[\rho,\beta\phi]$ via functional line integration
\eqref{EQGfunctionalViaFormalIntegral} and differentiating the result
with respect to $\beta \phi(r)$ gives very satisfactory agreement with
the test particle data, serving as a valuable consistency check. We
have performed this route, which is similar in spirit to the formal
functional method described by Evans~\cite{evans1992}, by using finite
numerical differentiation. The value of the scaled pair potential $\beta \phi(r)$ at a chosen distance $r$ is thereby increased by $d(\beta \phi) = 0.1$ and the resulting change in $F_\rmexc[\rho, \beta\phi]$ is monitored. Performing this finite differencing for all relevant values of $r$ yields the full functional derivative.

Figure \ref{FIGmetapairThreeRoutes} also shows results from the
numerical inversion of the homogeneous version of the Ornstein-Zernike
equation~\eqref{EQmetapairOrnsteinZernikeInhomogeneous} using as input
the bulk two-body direct correlation function obtained via the density
functional derivative \eqref{EQctwoFunctional} evaluated at constant
bulk density; see the bottom row of Fig.~\ref{FIGmetapairThreeRoutes}.
These results display quite pronounced oscillatory artifacts, which we
attribute to the well-known numerical intricacy of reliable
Ornstein-Zernike inversion rather than to a genuine shortcoming of the
neural functional. Also results from solving the corresponding linear
system of equations in real space are shown.

As further illustration of how the metachannel allows one to obtain
the above results, we display in
Fig.~\ref{FIGmetapairBulkMetacompressibility} representative results
for the scaled metafluctuation profile that forms the integrand in the
thermodynamic integral
\eqref{EQmetapiarGofrViaThermodynamicIntegration}.
Similarly to the one-body local compressibility $\chi_\mu(\rv)$, the
metacompressibility profile $\chi_\phi^b(r)$ measures local
fluctuations on the pair-correlation level. In test particle geometry,
$\chi_\phi^b(r)$ and $\chi_\mu(r)$ are hence closely connected,
cf.~also their respective general covariance definitions given in
\eqr{EQmetapairChiPhiDefinition} and above
\eqr{EQmetapairOZcompressibility}, which differ merely in the
replacement of $\beta N$ with $-\hat G(r)$.
The results shown in Fig.~\ref{FIGmetapairBulkMetacompressibility} are
for a bulk fluid system of Lennard-Jones-like particles, see the inset
in Fig.~\ref{FIGmetapairBulkMetacompressibility} for a depiction of
the form of $\beta\phi(r)$.  As the benchmark we have carried out
simulations, where sampling the covariance
\eqref{EQmetapairChiPhiBulk} provides a viable means of determining
$\chi_\phi^b(r)$. As a consistency check for the simulation reference,
the comparison to the result from the alternative route via the
thermodynamical derivative
\eqref{EQchiPhiBulkViaChemicalPotentialDerivative} yields excellent
confirmation.
The neural functional results are obtained via the bulk version
\eqref{EQmetaOrnsteinZernikeBulk} of the meta-Ornstein-Zernike
relation \eqref{EQmetaOrnsteinZernikeRelation} and we re-iterate that
all quantities on the right hand side follow from the same neural
metadensity functional. In particular the input that creates the
distance dependence is the metadirect correlation function
$c_\phi^b(r)$, as is available from the functional
derivative~\eqref{EQmetapairDefinitionCPhi} of
$c_1(x;[\rho,\beta\phi])$ with respect to $\beta\phi(r)$, evaluated at
constant bulk density.
While the results obtained from the original functional indicate
correct qualitative behaviour, they also display significant scatter
around the reference simulation data. In contrast, the regularized
functional suffers from no such artifacts and displays excellent
agreement with the simulation reference.

Fig.~\ref{FIGmetapairMetaDirectCorrelationFunctional} demonstrates the
accessibility of the fully resolved metadirect correlation functional
$c_\phi(x,r;[\rho,\beta\phi])$, we recall its definition
\eqref{EQmetapairDefinitionCPhi}, in a representative inhomogeneous
situation. The first panel depicts the highly inhomogeneous density
profile $\rho(x)$ that emerges in response to the imposed external
potential $V_\rmext(x)$ and we have switched to a repulsive Gaussian
interparticle potential to demonstrate the flexibility of the
metadensity functional. For comparison we show the result obtained via
the mean field approximation, $-\int dx'\rho(x)\delta(r-|x-x'|)$, see
the derivation given below \eqr{EQmetapairDefinitionCPhi}. While the
neural result shows qualitatively similar features, these are less
pronounced, which we deem to be more realistic. The result obtained
from the regularized functional shows reduced noise artifacts, but
also a slightly more pronounced overall variation. From our general
setup, we would expect this latter version to be the most accurate of
the three.

We show in Fig.~\ref{FIGmetapairTwoBodyDensity} results for the
two-body density and for the pair distance distribution $G(r)$ for a
representative inhomogeneous situation. The density profile is
sinusoidal and the particles interact again via Gaussian core
repulsion. Grand canonical Monte Carlo simulations provide reference
results for the two-body density distribution $\rho_2(x,x')$.  We have
ascertained that these results are consistent with the distance
statistics, see the caption of Fig.~\ref{FIGmetapairTwoBodyDensity}
for a detailed description.  Solution of the inhomogeneous
Ornstein-Zernike equation
\eqref{EQmetapairOrnsteinZernikeInhomogeneous}, where the neural
functional provides input for $c_2(x, x')$, yields results for the
two-body density that are in very good agreement with the simulation
data.

\section{Conclusions}
\label{SECconclusions}

In conclusion we have demonstrated that the metadensity functional
dependence on the form of the thermally scaled pair potential
$\beta\phi(r)$ provides fertile ground for deep and fresh insight into
both the formal structure of classical density functional theory and
the correlated equilibrium behaviour of fluids. We have restricted our
numerical work to one-dimensional systems that interact with
range-truncated interparticle forces. The theoretical metadensity
functional structure is general though and it holds, in principle, for
higher spatial dimensions and for long-ranged pair potentials.

Our present contribution complements Ref.~\citenum{kampa2024meta}, where
questions of Henderson inversion and inhomogeneous soft matter design were
addressed for one-dimensional systems. These topics were shown there to connect
naturally to the functional dependence on $\beta\phi(r)$~\cite{kampa2024meta}.
Neural density functionals, obtained via the local learning strategy, allow one
to capture accurately this `metadensity' functional dependence. In quite
striking contrast, this feat is not currently possible analytically, as there is
no generally applicable and reliable closed form for an excess free energy
functional known. A makeshift is the simple mean-field approximation, where the
dependence on $\phi(r)$ is linear, see the discussion below
\eqr{EQmetapairDefinitionCPhi} and the comparison with the neural metadensity
functional results presented in
Fig.~\ref{FIGmetapairMetaDirectCorrelationFunctional}. Tackling Henderson-type
inversion problems requires an accurate description of the functional dependence
on $\phi(r)$, which the neural metadensity functional can provide. In this
regard, it remains to be scrutinized precisely for which conditions $\phi(r)$
can be determined uniquely from structural data in inhomogeneous settings; the
neural metadensity functional can serve as valuable numerical guidance for this
task.

Crucially, our results demonstrate that the neural dependence on the
scaled pair potential $\beta\phi(r)$ transcends beyond being a mere
input `switch' of the neural network.  The `metachannel' rather
constitutes an accurate numerical representation of a formally
well-defined and unique functional dependence, which we have in
particular elucidated via Levy's constrained search method in
Sec.~\ref{SECmetadensityFunctionalDependence} within the presentation
of general concepts in Sec.~\ref{SECtheory}.  The data-driven evidence
for the validity of the present functional point of view stems from
the range of functional differentiation and functional line integral
relationships that we could exploit consistently and confirm to be
satisfied with high accuracy.

The present metadirect pair matching is to a certain extent inspired
by the machine learning of classical density functionals via pair
correlation matching as proposed by Dijkman {\it et
  al}~\cite{dijkman2024ml}. These authors base their method on
well-known statistical mechanical identities in density functional
formulation to constrain and train their neural network. Specifically,
the second density functional derivative of the excess free energy is
matched against corresponding results for {\it bulk} two-body direct
correlation functions. Importantly, Dijkman {\it et al.}~considered
the application of pair correlation matching to the truncated
Lennard-Jones fluid only, with no possibility at inference stage to
make predictions for other fluid types or for temperature values that
are different from that used for training.  Our present work expands
upon this scheme in two important ways. One is the incorporation of
training data for spatially {\it inhomogeneous} systems, which we deem
to be essential for reliable density functional learning; see the
extensive discussion given in Ref.~\citenum{sammueller2024pairmatching}.

The second difference of our present approach to bulk pair correlation
matching \cite{dijkman2024ml} is the depth at which the statistical
mechanical structure is put to work. We recall that for fixed
interparticle interaction potentials, thermal training
\cite{sammueller2024attraction, robitschko2025mixShort} of neural
functionals provides reliable access to temperature dependence and it
allows one to predict accurately structure, thermodynamics, and the
fluid phase behaviour across varying control parameters.  The
metadirect functional dependence generalizes the mere parametric
thermal dependence on $\beta$ to the functional dependence on $\beta
\phi(r)$, which incorporates both the (inverse) temperature parameter
$\beta$ as well as the form of the pair potential $\phi(r)$. As we
have laid out in Sec.~\ref{SECtheory} the metadensity functional
dependence allows one to put to work a range of classical formal
results given by Evans~\cite{evans1992}.

In particular we have constructed a regularized machine learning
scheme that is based on the pair correlation structure obtained from
functional differentiation with respect to the pair potential. We have
dubbed this the `metadirect' route and this is accessible only due to
the presence of the metachannel of the density functional and its
amenability to functional calculus; see the introduction of the
corresponding metadirect correlation functional in
Ref.~\citenum{kampa2024meta}.  We have described in detail in
Sec.~\ref{SECmetapairLearning} how the metadensity dependence
integrates itself into the local density functional learning and that
a two-stage machine learning method forms an efficient and effective
regularization scheme. We have exploited that these deep statistical
mechanical roots can be integrated straightforwardly within modern
computing paradigms of supervised neural network training and
automatic differentiation. Together with fast neural functional line
integration wide exploration and exploitation of the functional
formulation is possible. It is in this sense that we put a variant of
physics-informed machine learning to work via incorporation of
statistical mechanical first principles. Although the supervised
learning aspect remains based on simulation data, our present strategy
differs from more common data-driven approaches that seek to represent
a desired physical target quantity straight away by a surrogate model,
see e.g.\ for a topical example the recent study by Liebchen and
co-workers to learn the microstructure in active
matter~\cite{dasgupta2026}.

As an outlook on potential future work, we mention several interesting
topics. Instead of generating second-stage training data for
metadirect pair matching self-consistently via the neural test
particle procedure, one could alternatively use Monte Carlo results
for the bulk metacompressibility profile. Albeit being more costly to
obtain than the neural density functional results, this method could
potentially avoid any latent `passing down' of inaccuracies from the
first-stage neural functional.  Related to this, how the bulk method
by Dijkman {\it et al.}~\cite{dijkman2024ml} would fare if generalized
to include meta-dependence on the pair potential is an interesting
question.  Although only bulk states remain probed, incorporating the
metadensity functional dependence should yield additional beneficial
constraints on the resulting neural functional.

Interesting work could be aimed at elucidating the relationship of the
bulk metacompressibility with the recent `3g'-sum rule
\cite{sammueller2023whatIsLiquid} that connects $g(r)$ to two further
(force-force and force-gradient) pair correlation functions. More
generally, statistical mechanical gauge invariance
\cite{hermann2021noether, mueller2024gauge, mueller2024whygauge}
provides a powerful toolbox of formally exact sum rules that carry
significant potential for the development of further physics-informed
regularization schemes and for imposing concrete constraints on neural
functionals.

While the results presented here attest to the prowess of the metadensity
functional concept \cite{kampa2024meta}, much important work remains to be
addressed. All neural investigations in the present work concern one-dimensional
systems of particles that interact via short-ranged (truncated) pair potentials.
The metadensity dependence is only available within the finite input window size
of $\beta\phi(r)$. Going beyond these limitations provides much fertile ground
for future work. We believe that in particular taming spherically symmetric
situations, which are a requirement for carrying out test particle calculations
in $d>1$, will be an important step to shed further light on the
interrelationship of statistical mechanics and geometry. In this context
relevant questions concerning the behaviour at curved substrates, such as fluid
adsorption and solvation effects, could be addressed from an arguably new angle.
Our framework also ties in well with modern questions regarding the behaviour of
responsive colloids that adapt their interparticle interaction potential in
response to various different stimuli and physical situations
\cite{monchojorda2020, bley2021, bley2022, baul2021, lopezmolina2024,
  monchojorda2023}. Addressing further generalizations to practically relevant
types of soft matter, in particular to systems involving long-ranged
interactions that arise e.g.\ in charged or dipolar fluids,
\cite{bui2024neuralrpm,bui2025dielectrocapillarity}
constitutes valuable future work.

\bigskip

{\bf Acknowledgments.} We thank Bob Evans for much inspiration to
carry out this research, him and Jacco Dijkman for useful discussions,
and the Special Issue Editors of J.\ Phys.\ Chem.\ B for their
patience in receiving this article. This work is supported by the DFG
(Deutsche Forschungsgemeinschaft) under project no.~551294732.

\bigskip

{\bf Data availability}. The data that support the findings of this
article are openly available \cite{kampa2026pairmatchingData}.


\end{document}